\documentclass[twocolumn,showpacs,amsmath,amssymb,aps,prd]{revtex4}

\usepackage{bm}

\newcommand{\itGamma}{{\mathit{\Gamma}}}
\newcommand{\del}{\partial}

\newcommand{\cM}{{\cal M}}

\begin{document}

\title{Classical String in Curved Backgrounds}

\author{Milovan Vasili\'c}
 \email{mvasilic@phy.bg.ac.yu}
\author{Marko Vojinovi\'c}
 \email{vmarko@phy.bg.ac.yu}
\affiliation{Institute of Physics, P.O.Box 57, 11001 Belgrade, Serbia}

\date{\today}

\begin{abstract}
The Mathisson-Papapetrou method is originally used for derivation of the particle
world line equation from the covariant conservation of its stress-energy tensor. We
generalize this method to extended objects, such as a string. Without specifying the
type of matter the string is made of, we obtain both the equations of motion and
boundary conditions of the string. The world sheet equations turn out to be more
general than the familiar minimal surface equations. In particular, they depend on
the internal structure of the string. The relevant cases are classified by examining
canonical forms of the effective $2$-dimensional stress-energy tensor. The case of
homogeneously distributed matter with the tension that equals its mass density is
shown to define the familiar Nambu-Goto dynamics. The other three cases include
physically relevant massive and massless strings, and unphysical tahyonic strings.
\end{abstract}


\maketitle

\section{\label{IntroductionSection}Introduction}

The original motivation for introducing strings in particle
physics came from the analysis of meson resonances. As it appears,
the known resonances, characterized by the angular momentum $J$
and the mass $M$, follow the pattern $J=\alpha M^2 + {\rm const.}$,
where $\alpha$ is a universal constant. These are called
Regge trajectories.

To explain Regge trajectories, the meson resonances are viewed as excited
$2$-quark bound states. It has been shown then that relativistic rotating
string with light quarks attached to its ends indeed reproduces the above
pattern. The string is characterized by the tension alone, and has no other
structure. It was realized later that realistic field configurations with
such properties really exist. Such is, for example, the flux tube solution
of Ref. \cite{Nielsen1973}.

In what follows, we shall not be concerned with particular field-theoretical
models that accommodate flux tubes, or any other linelike configurations of
fields. We shall merely assume that such kink configurations exist, and try
to draw from it as much information as possible. In particular, we want to
obtain the world sheet equations of motion.

Our motivation for considering stringy shaped matter in curved backgrounds
is twofold. First, as we have already explained, realistic strings (like
flux tubes) are really believed to exist, and to be relevant for the
description of hadronic matter. Second, the basic Nambu-Goto string action
\cite{Nambu, Goto} is in literature often modified to include interaction
with additional background fields. Apart from the target-space metric, the
anti-symmetric tensor field $B_{\mu\nu}(x)$ and the dilaton field $\Phi(x)$
are considered \cite{B1, B2, B3, B4}. While spacetime metric has obvious
geometric interpretation, the background fields $B_{\mu\nu}(x)$ and
$\Phi(x)$ do not. The attempts have been made in literature to interpret
$B_{\mu\nu}$ and $\Phi$ as originating from the background torsion and
nonmetricity, respectively \cite{G1, G2, G3, G4, G5}. It seems to us that
string dynamics in target-spaces of general geometry is worth considering.

Basically, we are interested in the influence of the target-space torsion on
the string dynamics. Our idea is to consider a field-theoretical model that
naturally includes torsion (like Poincar\' e gauge theory of gravity) and
find the equations of motion of a stringy shaped material object. Hopefully,
the effective action of Refs. \cite{B1, B2, B3, B4} would be recovered,
and the real geometric nature of the background fields $B_{\mu\nu}(x)$ and
$\Phi(x)$ found.

In this paper, we shall restrict our considerations to the simpler case of
purely Riemannian spacetime. Thus, the geometry is given in terms of the
metric tensor alone, and the dynamics is governed by the Einstein's
equations
\begin{equation} \label{AjnstajnoveJednacine}
R_{\mu\nu}-\frac{1}{2}g_{\mu\nu}R = 8\pi G \,T_{\mu\nu}.
\end{equation}
The stress-energy tensor of matter fields is symmetric,
$T^{\mu\nu}=T^{\nu\mu}$, and covariantly conserved,
$\nabla_{\nu}T^{\mu\nu}=0$. In Riemannian spacetimes, the connection in the
definition of the covariant derivative
$\nabla_{\nu}v^{\mu}=\del_{\nu}v^{\mu}+\itGamma^{\mu}{}_{\rho\nu}v^{\rho}$
is the Levi-Civita connection. As a consequence, the stress-energy covariant
conservation law $\nabla_{\nu}T^{\mu\nu}=0$ is rewritten in the form
\begin{equation} \label{ZOTEIuZgodnomObliku}
\del_{\nu}(\sqrt{-g}T^{\mu\nu})+{\itGamma^{\mu}}_{\rho\nu}
\sqrt{-g}T^{\rho\nu} = 0 .
\end{equation}
This equation will be the starting point in our analysis of motion of
extending objects in curved spacetimes. For practical purposes, we shall
consider extended objects with the atributes of test bodies. This way, their
influence on the background geometry becomes negligible.

The method we use is a straightforward generalization of the
Mathisson-Papapetrou method for pointlike matter \cite{Mathison,
Papapetrou1951}. It boils down to the analysis of the covariantly conserved
stress-energy tensor of matter fields, without specifying their nature. The
basic assumption used is the existence of a stringlike localized kink
solution in a curved background. Then, the world sheet effective equations
of motion are obtained in the approximation of an infinitely thin string.

Although the case under consideration is a simple one (torsionless,
Riemannian geometry), the resulting world sheet equations turn out to depend
on the internal structure of the string. This dependence enters the
equations of motion through the effective $2$-dimensional stress-energy
tensor of the string. Classifying possible canonical forms of the
stress-energy $2$-tensor, we shall discover a specific distribution of
matter characterized by the tension alone. Its world sheet dynamics is fully
determined by the target-space geometry, and coincides with that of the
known Nambu-Goto string \cite{Nambu, Goto}. Different matter distributions
will contribute to different world sheet equations, and different boundary
conditions. In particular, if the string mass is localized in a point, the
world sheet turns into a conventional geodesic line.

The layout of the paper is as follows. In Sec.\ \ref{ParticleSection}, the
point particle is considered as a demonstration of our method and
conventions. The known result is reproduced, but the emphasis is put on the
fact that the mass parameter transforms as $1$-dimensional stress-energy
tensor. In Sec.\ \ref{StringSection}, the effective world sheet equations
are derived from the covariant conservation law of the stress-energy tensor
of matter fields. Instead of the mass parameter in the point particle case,
here, the effective $2$-dimensional stress-energy tensor $m^{ab}$ appears to
characterize the internal structure of the string. The world sheet equations
imply the covariant conservation of $m^{ab}$ with respect to the induced
$2$-dimensional world sheet metric $\gamma_{ab}$. If the string is open, the
world sheet equations also include some boundary conditions. Sec.\
\ref{ExamplesSection} is devoted to the analysis of possible canonical forms
of $m^{ab}$, and to some examples. It is shown how the assumption of
homogeneously distributed matter with the tension that equals its mass
density leads to the known Nambu-Goto string. As an example, the
Nielsen-Olesen vortex line solution of a Higgs type scalar electrodynamics
is briefly examined \cite{Nielsen1973}. In Sec.\ \ref{Conclusions} we give
our final remarks. In particular, we emphasize that the results we have
obtained are easily generalized to hold for any $p$-brane in an external
Riemannian spacetime.

Our conventions are as follows. Greek indices from the middle of the
alphabet, $\mu,\nu,\dots$, are the target-space indices, and run over
$0,1,\dots,D-1$. Greek indices from the beginning of the alphabet,
$\alpha,\beta,\dots$, refer to the spatial section of the target-space, and
run over $1,2,\dots,D-1$. Latin indices $a,b,\dots$ are the world sheet
indices and run over $0,1$. The target-space and world sheet coordinates are
denoted by $x^{\mu}$ and $\xi^{a}$, respectively. The target-space and
world sheet metric tensors are denoted by $g_{\mu\nu}(x)$ and
$\gamma_{ab}(\xi)$, respectively. The signature convention is defined by
$\eta_{\mu\nu}=\text{diag}(1,-1,\dots,-1)$,
$\eta_{ab}=\text{diag}(1,-1)$. Target-space indices $\mu,\nu,\dots$ are
lowered and raised by the target-space metric $g_{\mu\nu}$ and its inverse
$g^{\mu\nu}$, while world sheet indices $a,b,\dots$ are lowered and raised
by the world sheet metric $\gamma_{ab}$ and its inverse $\gamma^{ab}$.
Throughout the paper, we shall restrict our considerations to $D=4$, but we
note that all the results remain valid for arbitrary spacetime dimension.

\section{\label{ParticleSection}Particle dynamics}

We begin with the treatment of a point particle in a curved background
spacetime. The problem was studied in the early days of relativity by
Einstein, Infeld, Hoffman, Mathisson, Papapetrou and others \cite{Mathison,
Papapetrou1951, Pryce, Tulczyjew, Weyssenhoff, Dixon}. Here, we formalize
the calculations, and adjust the algorithm for the case of a string in the
next section.

\subsection{Stress-energy tensor}

We need a general form of the stress-energy tensor, suitable for the
description of a point particle.

Let us introduce a timelike curve $x^{\mu}=z^{\mu}(\tau)$ in spacetime,
with $\tau$ an arbitrary parameter. We shall consider spacetimes of topology
$\Sigma \times R$, where $R$ stands for time, and $\Sigma$ is an arbitrary
$3$-manifold representing spatial sections. Both, the spacetime and the
curve are supposed to be {\it non-degenerate} and {\it complete}. In simple
terms, only infinite curves are considered. This way, the unphysical matter
distributions, such as instantons, are excluded.

In general, the spacetime coordinates $x^{\mu}$, and the world line
parameter $\tau$ are arbitrary. Still, we shall partially fix this freedom
to make the exposition more transparent. First, the coordinates $x^{\mu}$
are chosen in accordance with the demand that equal-time surfaces are
spacelike. As a consequence, our curve intersects these surfaces only once,
and the function $z^0(\tau)$ becomes invertible. Second, the time coordinate
$x^0\equiv t$ is chosen to parametrize the curve. The choice $\tau = t$, or
equivalently, $z^0(\tau)\equiv\tau$, puts the constraint $u^0=1$ to the form
of the tangent vector $u^{\mu}\equiv dz^{\mu}/d\tau$. Otherwise, it is an
arbitrary time-like vector satisfying $g_{\mu\nu}u^{\mu}u^{\nu}\equiv u^2 >
0$.

Next, we expand $\sqrt{-g}T^{\mu\nu}(x)$ into the $\delta$ function series
around the point $\vec{x}=\vec{z}(t)$, while treating the time coordinate as
a parameter. Using the formula (\ref{app:RazvojFunkcijePoDeltaFunkciji})
(case $d=3$), we have:
\begin{eqnarray*}
\sqrt{-g}T^{\mu\nu}(t,\vec{x})
&=& \sqrt{\gamma}b^{\mu\nu}(t)\delta^{(3)}(\vec{x}-\vec{z}(t))\nonumber\\
& & {}+\sqrt{\gamma}b^{\mu\nu\alpha}(t)\del_{\alpha}\delta^{(3)}
      (\vec{x}-\vec{z}(t))+\cdots .
\end{eqnarray*}
Here, $\gamma$ is the induced metric on the curve. It is defined by
$ds^2=g_{\mu\nu}(z)dz^{\mu}dz^{\nu}\equiv \gamma(\tau)d\tau^2$, and is
introduced for later convenience. The coefficients in the expansion are
given by the formula (\ref{app:KoeficijentiRazvojaFunkcijePoDeltaFunkciji}):
\begin{eqnarray*}
&& \sqrt{\gamma}b^{\mu\nu}=
   \int d^3\vec{x}\sqrt{-g}T^{\mu\nu},           \\
&& \sqrt{\gamma}b^{\mu\nu\alpha}=-\int d^3\vec{x}
   (x^{\alpha}-z^{\alpha})\sqrt{-g}T^{\mu\nu},
\end{eqnarray*}
etc. Note that the coefficients in the expansion equal the Papapetrou
moments $M^{\mu\nu}, M^{\alpha\mu\nu}$ etc., as defined in
\cite{Papapetrou1951}.

Now, we introduce the basic assumption about matter. It is localized around
the line $z^{\mu}(\tau)$, ie. the {\it stress-energy tensor drops
exponentially to zero as we move away from the line}. Of course, this
assumption means that field equations allow solutions of such a type. Field
theories that have such properties are known to exist, but we shall not be
interested in details of particular models.

As a consequence of this assumption, each coefficient $b^{\mu\nu\alpha_1
\dots \alpha_n}$ gets smaller as $n$ gets larger. In the lowest
approximation (the so called {\it single pole} approximation), all $b$-s
except the first are neglected, and we end up with
\begin{equation} \label{TEIuSiglePoleAproksimaciji}
\sqrt{-g}T^{\mu\nu}(t,\vec{x})=\sqrt{\gamma}b^{\mu\nu}(t)
\delta^{(3)}(\vec{x}-\vec{z}(t)).
\end{equation}
This equation is not covariant with respect to the target-space coordinate
transformations. To cast it into a covariant form, we add an extra $\delta$
function and an extra integration. Thus, we obtain
\begin{equation} \label{ParticleCovariantTEI}
\sqrt{-g}T^{\mu\nu}(x)=\int d\tau\sqrt{\gamma}b^{\mu\nu}(\tau)
\delta^{(4)} (x - z(\tau)),
\end{equation}
which reduces to (\ref{TEIuSiglePoleAproksimaciji}) in the gauge
$z^0(\tau)=\tau$. This form of stress-energy tensor is covariant with
respect to both spacetime coordinate transformations and world line
reparametrizations. From the known transformation properties of
$T^{\mu\nu}$, we infer the transformation properties of $b^{\mu\nu}$. It is
a tensor with respect to general coordinate transformations, and scalar with
respect to world line reparametrizations. Equation
(\ref{ParticleCovariantTEI}) describes matter localized around the line
$z^{\mu}(\tau)$, and in this form, we shall use it to solve the Eq.
(\ref{ZOTEIuZgodnomObliku}).

\subsection{Equations of motion}

We look for the solution of the Eq. (\ref{ZOTEIuZgodnomObliku}) in the
form (\ref{ParticleCovariantTEI}), where $b^{\mu\nu}(\tau)$ and
$z^{\mu}(\tau)$ are the unknown functions to be determined. Thus, we
obtain
$$
\int d\tau \sqrt{\gamma} \left[ b^{\mu\nu} \del_{\nu} \delta^{(4)}(x-z) +
b^{\rho\nu} {\itGamma^{\mu}}_{\rho\nu} \delta^{(4)}(x-z) \right]=0 .
$$
Viewed as the $\delta$ function expansion, this equation is decoupled into a
pair of equations determining $z^{\mu}$ and $b^{\mu\nu}$. To see this, we
multiply the equation with an arbitrary function $f(x)$ of {\it compact
support}, and integrate over the spacetime. The compact support of $f(x)$
allows switching the order of integrations, which results in
\begin{equation} \label{JednacinaPomnozenaProizvoljnomFunkcijom}
\int d\tau\sqrt{\gamma}\left[-b^{\mu\nu}\frac{\del f(z)}{\del z^{\nu}}
 + b^{\rho\nu} {\itGamma^{\mu}}_{\rho\nu}(z) f(z)\right] = 0 .
\end{equation}
The scalar field $f(x)$ is arbitrary, but the longitudinal component of the
gradient $ f_{,\nu}\equiv\del f(z)/\del z^{\nu}$ is not independent of
$f(z)$. So, we decompose the gradient $f_{,\nu}$ into the parallel and
orthogonal components:
$$
f_{,\nu} = f_{\nu}^{\perp} + f^{\parallel}u_{\nu}.
$$
By definition, $f_{\nu}^{\perp}u^{\nu}=0$, and the coefficient
$f^{\parallel}$ is obtained from the identity $df/d\tau\equiv f_{,\nu}
u^{\nu}$. Thus,
$$
f_{,\nu} = f_{\nu}^{\perp} + \frac{1}{\gamma} \frac{df}{d\tau} u_{\nu}.
$$
Now, $f(z(\tau))$ and $f_{\nu}^{\perp}(z(\tau))$ are independent and
arbitrary. We arrange (\ref{JednacinaPomnozenaProizvoljnomFunkcijom}) as
follows:
\begin{widetext}
\begin{equation} \label{ZgodanOblikZaAnalizu}
\int d\tau \left\{
\sqrt{\gamma} b^{\mu\nu} f_{\nu}^{\perp} -
\left[\frac{d}{d\tau}\left(\frac{1}{\sqrt{\gamma}}u_{\nu}b^{\mu\nu}\right)
+\sqrt{\gamma}b^{\rho\nu}{\itGamma^{\mu}}_{\rho\nu}\right]f +
\frac{d}{d\tau}\left(\frac{1}{\sqrt{\gamma}}u_{\nu}b^{\mu\nu}f\right)
\right\} =0 .
\end{equation}
\end{widetext}
Each of the three terms in the above equation must separately vanish.

The third term gives no contribution to the world line equations. Indeed,
the world line is assumed to be infinite, and the function $f$ to have
compact support. As a consequence, the corresponding boundary integral
vanishes.

The second term gives the world line equations in the form
\begin{equation} \label{PrvaDekuplovanaJednacina}
\frac{d}{d\tau} \left( \frac{1}{\sqrt{\gamma}} u_{\nu}b^{\mu\nu} \right)
+ \sqrt{\gamma} b^{\rho\nu} {\itGamma^{\mu}}_{\rho\nu} =0 .
\end{equation}
Viewed as an equation for $z^{\mu}(\tau)$, it contains the undetermined
coefficients $b^{\mu\nu}(\tau)$.

In the first term, we decompose $b^{\mu\nu}$ into the parallel and
orthogonal components with respect to the second index:
$$
b^{\mu\nu} = b^{\mu\nu}_{\perp} + b^{\mu} u^{\nu},
$$
where $b^{\mu\nu}_{\perp}u_{\nu}\equiv 0$. Then, vanishing of the term
$b^{\mu\nu}f_{\nu}^{\perp}$ for every $f_{\nu}^{\perp}$ implies
$b^{\mu\nu}_{\perp}=0$, and consequently,
$$
b^{\mu\nu} = b^{\mu}u^{\nu}.
$$
As $b^{\mu\nu}$ is a symmetric tensor, $b^{\mu}u^{\nu}$ must equal
$b^{\nu}u^{\mu}$, so that $b^{\mu} \propto u^{\mu}$. Therefore,
\begin{equation} \label{DrugaDekuplovanaJednacina}
b^{\mu\nu} = m u^{\mu}u^{\nu},
\end{equation}
where $m(\tau)$ is an arbitrary coefficient. We see that, up to a
multiplicative term, $b^{\mu\nu}$ is fully determined by $z^{\mu}(\tau)$.

We can now substitute (\ref{DrugaDekuplovanaJednacina}) into
(\ref{PrvaDekuplovanaJednacina}) and obtain
\begin{equation} \label{ResenaPrvaDekuplovanaJednacina}
\frac{d}{d\tau}\left(\sqrt{\gamma} m u^{\mu}\right)
+\sqrt{\gamma} m {\itGamma^{\mu}}_{\rho\nu}u^{\rho}u^{\nu} = 0 .
\end{equation}
This is our final, covariant world line equation. It contains the
undetermined $m(\tau)$, but this coefficient is constrained by the very same
equation. Indeed, the projection of (\ref{ResenaPrvaDekuplovanaJednacina})
on the tangent vector $u^{\mu}$ can straightforwardly be brought to the
simple form
\begin{equation} \label{KovarijantnoOdrzanjeMzaCesticu}
\frac{d}{d\tau} \left( m\gamma \right) = 0 .
\end{equation}
We see that $m\gamma$ is a constant of motion, and consequently, it can
easily be eliminated from the world line equations. In fact, using the
proper distance $s$ to parametrize the curve (which is equivalent to fixing
the gauge $\gamma=1$), we get $m={\rm const.}$, and restore the standard
geodesic equation
$$
\frac{d^2 z^{\mu}}{ds^2}+{\itGamma^{\mu}}_{\rho\nu}\frac{dz^{\rho}}{ds}
\frac{dz^{\nu}}{ds} = 0 .
$$

\subsection{Discussion}

The world line equations we have obtained are manifestly covariant with
respect to both general coordinate transformations and world line
reparametrizations. The quantities $b^{\mu\nu}$, $u^{\mu}$, and $m$, beside
being spacetime tensors (second rank tensor, vector, and scalar,
respectively), are also tensors with respect to the reparametrizations $\tau'
= \tau'(\tau)$ (scalar, vector, and second rank tensor, respectively).
In particular,
$$
m'(\tau') = \frac{d\tau'}{d\tau} \frac{d\tau'}{d\tau} m(\tau) .
$$
Thus, $m$ transforms as a second rank contravariant tensor with respect to
world line reparametrizations. This gives us the idea that $m$ can be viewed
as an effective one-dimensional stress-energy tensor of the pointlike
matter. In support of this interpretation, note that the earlier established
conservation of $m\gamma$, as given by
(\ref{KovarijantnoOdrzanjeMzaCesticu}), can be rewritten as
\begin{equation}\label{CovariantlyConservedMass}
\nabla_{\tau}m = 0 ,
\end{equation}
where $\nabla_{\tau}$ stands for the one-dimensional, Riemannian covariant
derivative ($\nabla_{\tau}v\equiv \del_{\tau}v+\itGamma v $, where
$\itGamma$ is one-dimensional Levi-Civita connection). Thus, our coefficient
$m$ can really be viewed as an effective, {\it covariantly conserved}
one-dimensional energy-momentum tensor. In this respect, $m$ should be
considered the particle mass.

The results of this section can be summarized as follows. The stress-energy
conservation Eq. (\ref{ZOTEIuZgodnomObliku}) is applied to a linelike
distribution of matter. In the lowest approximation, such matter
distribution is covariantly described by (\ref{ParticleCovariantTEI}), with
$z^{\mu}(\tau)$ and $b^{\mu\nu}(\tau)$ the unknown functions. We found that
(a) the world line, parametrized by the proper distance $s$, satisfies the
geodesic equation
$$
\frac{d^2 z^{\mu}}{ds^2}+{\itGamma^{\mu}}_{\rho\nu}
\frac{dz^{\rho}}{ds} \frac{dz^{\nu}}{ds}=0 ,
$$
and (b) the stress-energy tensor takes the form
$$
\sqrt{-g}T^{\mu\nu}(x)=m\int ds u^{\mu}u^{\nu}\delta^{(4)}(x-z(s)),
$$
with $m$ a constant interpreted as the particle mass.

The above results are obtained in the lowest approximation in the $\delta$
function expansion. If, however, the second term ({\it pole-dipole}
approximation), or higher order terms were included, the world line equation
would depend on the internal structure of the particle. In particular, the
particle angular momentum would couple to the spacetime curvature, giving
deviations from the geodesic trajectory. The analysis of the higher order
particle moments has extensively been done in the literature (see, for
example \cite{Papapetrou1951}). Here, we just prepare the setting for the
study of string dynamics in the next section.

\section{\label{StringSection}String dynamics}

The calculations presented in the previous section are well known, and there
are papers \cite{Yasskin1980, Nomura1991} which generalize the procedure to
include torsion, and explore modifications that it brings to the theory.
However, this research has been focused on the particle case, and we want to
address the problem of finding equations of motion of an extended object,
such as string. In this section, we generalize the Papapetrou method to
linelike matter, and present the results.

\subsection{Stress-energy tensor}

As in the particle case, we begin with the stress-energy covariant
conservation law in the form (\ref{ZOTEIuZgodnomObliku}).

In contrast to the particle, the string is an extended, one-dimensional
object whose trajectory is not a world line, but rather a two-dimensional
world sheet $\cM$. Let us introduce a two-dimensional surface
$x^{\mu}=z^{\mu}(\xi^a)$ in spacetime, where $\xi^0$ and $\xi^1$ are the
surface coordinates. We shall assume that the surface is {\it everywhere
regular}, and the coordinates $\xi^a$ well defined. As in the particle case,
we shall consider only time-infinite string trajectories. This means that
every spatial section of the spacetime has nonempty intersection with the
world sheet. As for the intersection itself, it is supposed to be of finite
length. Thus, only closed, or finite open strings are considered. In the
conventional parametrization, $\xi^0\equiv\tau$ goes from minus to plus
infinity, while $\xi^1\equiv\sigma$ takes values in the interval $[0,\pi]$.
In this parametrization, the world sheet boundary is defined by the
coordinate lines $\sigma=0$ and $\sigma=\pi$.

In what follows, we shall frequently use the notion of the world sheet
coordinate vectors
$$
u^{\mu}_a \equiv \frac{\del z^{\mu}}{\del \xi^a},
$$
and the world sheet induced metric tensor
$$
\gamma_{ab} = g_{\mu\nu} u^{\mu}_a u^{\nu}_b .
$$
If the world sheet is regular, and the coordinates $\xi^a$ are well defined,
the two tangent vectors $u^{\mu}_0$ and $u^{\mu}_1$ are linearly
independent. The induced metric is assumed to be nondegenerate, $\det
(\gamma_{ab}) \neq 0$, and of Minkowski signature $(+,-)$. With this
assumption, each point on the world sheet accommodates a timelike tangent
vector. This is how the notion of the timelike curve is generalized to the
two-dimensional case. In what follows, we shall also discuss the situations
in which this assumption is violated on the world sheet boundary.

We shall restrict our considerations to $4$-dimensional spacetimes. As
before, we expand the stress-energy tensor into a $\delta$ function series
around the world sheet. The procedure is basically the same as in the
particle case, the only difference being in the use of $\delta^{(2)}$
instead of $\delta^{(3)}$ functions. In the single-pole approximation, we
drop all the terms in the expansion except the leading one. In this
approximation, the stress-energy tensor contains no $\delta$ function
derivatives. Similar to the particle case, the covariantization is achieved
by employing two more $\delta$ functions, and two more integrations. Thus,
we obtain a covariant expression for the stress-energy tensor:
\begin{equation} \label{StringCovariantTEI}
\sqrt{-g}T^{\mu\nu}(x)=\int d^2\xi\sqrt{-\gamma}b^{\mu\nu}(\xi)
\delta^{(4)}(x-z(\xi)).
\end{equation}
The coefficients $b^{\mu\nu}$ transform covariantly with respect to both
target-space and world sheet reparametrizations.

\subsection{Equations of motion}

Using the ansatz (\ref{StringCovariantTEI}) in the Eqs.
(\ref{ZOTEIuZgodnomObliku}) yields
$$
\int d^2\xi \sqrt{-\gamma} \left[ b^{\mu\nu} \del_{\nu}
\delta^{(4)}(x\!-\!z) + b^{\rho\nu} {\itGamma^{\mu}}_{\rho\nu}
\delta^{(4)}(x\!-\!z) \right]=0 .
$$
The left-hand side of this equation is almost in the form of the $\delta$
function series. To make use of this, we multiply the equation with an
arbitrary function $f(x)$ of {\it compact support}, and integrate over the
spacetime. The compact support of $f(x)$ allows switching the order of
integrations. Using partial integration in the first term, we get
\begin{equation} \label{StrunskaJednacinaPomnozenaProizvoljnomFunkcijom}
\int d^2\xi\sqrt{-\gamma}\left[-b^{\mu\nu}\frac{\del f(z)}
{\del z^{\nu}}+b^{\rho\nu}{\itGamma^{\mu}}_{\rho\nu}(z) f(z)\right]=0 .
\end{equation}
The scalar field $f(x)$ is arbitrary, but the projection of the gradient
$f_{,\nu}\equiv \del f(z)/\del z^{\nu}$ on the world sheet is not
independent of $f(z)$. So, we decompose the gradient $f_{,\nu}$ into the
parallel and orthogonal components:
$$
f_{,\nu} = f_{\nu}^{\perp} + f_a^{\parallel}u^a_{\nu}.
$$
By definition, $f_{\nu}^{\perp}u^{\nu}_a=0$, and the coefficients
$f_a^{\parallel}$ can be expressed through $\del f/\del\xi^a \equiv
f_{,\nu}u^{\nu}_a$. Thus, we have:
$$
f_{,\nu} = f_{\nu}^{\perp} + \frac{\del f}{\del \xi^a} u_{\nu}^a .
$$
Now, $f$ and $f_{\nu}^{\perp}$ are mutually independent on the world sheet.
We arrange (\ref{StrunskaJednacinaPomnozenaProizvoljnomFunkcijom}) as
follows:
\begin{widetext}
\begin{equation} \label{StrunskiZgodanOblikZaAnalizu}
\int_{\cM}d^2\xi \left\{
\sqrt{-\gamma} b^{\mu\nu} f_{\nu}^{\perp} -
\left[\frac{\del}{\del\xi^a}\left(\sqrt{-\gamma}b^{\mu\nu}u_{\nu}^a\right)
+ \sqrt{-\gamma} b^{\rho\nu} {\itGamma^{\mu}}_{\rho\nu} \right] f +
\frac{\del}{\del\xi^a}\left(\sqrt{-\gamma}b^{\mu\nu}u_{\nu}^a f\right)
\right\} = 0 .
\end{equation}
\end{widetext}
Owing to the arbitrariness and mutual independence of $f$ and
$f_{\nu}^{\perp}$, each of the three terms in the integrand of the above
equation must separately vanish.

The third term is a two-divergence, and by Stokes theorem, reduces to a line
integral over the boundary $\del\cM$. It must vanish for any choice of the
function $f$ evaluated on the boundary, and therefore, implies the boundary
condition
\begin{equation} \label{GranicniUsloviStrune}
\sqrt{-\gamma}b^{\mu\nu}u_{\nu}^a n_a \Big|_{\del\cM} =0 .
\end{equation}
Here, $n_a$ is the outward directed normal to the boundary $\del\cM$. If the
boundary line $\xi^a=\xi^a(\varsigma)$ is parametrized by some parameter
$\varsigma$, the normal will take the form
\begin{equation} \label{DefinicijaNormaleNaGranicu}
n_a = \varepsilon_{ab}\frac{d\xi^b}{d\varsigma},
\end{equation}
where $\varepsilon_{ab}$ is the antisymmetric Levi-Civita tensor. The
boundary conditions (\ref{GranicniUsloviStrune}) do not appear if the string
is closed. In that case, $\del\cM=\emptyset$, and the third term of Eq.
(\ref{StrunskiZgodanOblikZaAnalizu}) identically vanishes.

The vanishing of the second term in (\ref{StrunskiZgodanOblikZaAnalizu})
yields the world sheet equation
\begin{equation} \label{PrvaDekuplovanaJednacinaStrune}
\frac{\del}{\del\xi^a}\left(\sqrt{-\gamma}b^{\mu\nu}u_{\nu}^a\right)
+ \sqrt{-\gamma}b^{\rho\nu}{\itGamma^{\mu}}_{\rho\nu}=0 .
\end{equation}
It generalizes the particle world line equation
(\ref{PrvaDekuplovanaJednacina}). As an equation for $z^{\mu}(\xi)$, it
should be supplemented by appropriate constraints on the unknown
coefficients $b^{\mu\nu}$.

Finally, consider the first term. First, we split $b^{\mu\nu}$ into the
parallel and orthogonal components with respect to the second index:
$$
b^{\mu\nu}=b^{\mu\nu}_{\perp}+b^{\mu a} u^{\nu}_a ,
$$
where $b^{\mu\nu}_{\perp}u_{a\nu}\equiv 0$. Vanishing of the term
$b^{\mu\nu}f_{\nu}^{\perp}$ for every $f_{\nu}^{\perp}$ implies
$b^{\mu\nu}_{\perp}=0$, and we are left with
$$
b^{\mu\nu}=b^{\mu a}u^{\nu}_a .
$$
Again, we make use of the symmetry of the stress-energy tensor, and obtain
$$
b^{\mu a}u^{\nu}_a=b^{\nu a}u^{\mu}_a .
$$
This means that $b^{\mu a}$  is a linear combination of vectors
$u^{\mu}_a$, and consequently,
\begin{equation} \label{ResenaDrugaDekuplovanaJednacinaStrune}
b^{\mu\nu} = m^{ab} u^{\mu}_a u^{\nu}_b .
\end{equation}
Here, $m^{ab}(\xi)$ are arbitrary coefficients. They transform as scalars
with respect to spacetime diffeomorphisms, and as components of a
contravariant symmetric second rank tensor with respect to the world sheet
reparametrizations. Apart from arbitrariness in $m^{ab}$, $b^{\mu\nu}$ is
fully determined by $u^{\mu}_a$, which are, in turn, fully determined by
$z^{\mu}(\xi)$. The boundary conditions (\ref{GranicniUsloviStrune}) now
read
\begin{equation} \label{FinalBoundaryConditions}
\sqrt{-\gamma}m^{ab}n_b u^{\mu}_a\Big|_{\del\cM} =0 ,
\end{equation}
while the world sheet equation (\ref {PrvaDekuplovanaJednacinaStrune}) takes
the form
\begin{equation} \label{ResenaPrvaDekuplovanaJednacinaStrune}
\del_a \left( \sqrt{-\gamma} m^{ab} u^{\mu}_b \right)+
\sqrt{-\gamma}m^{ab}u^{\rho}_a u^{\nu}_b{\itGamma^{\mu}}_{\rho\nu} =0 .
\end{equation}

The world sheet equations can be written in a manifestly covariant way. To
this end, we make use of the {\it total covariant derivative} $\nabla_a$,
which acts on both spacetime and world sheet indices:
$$
\nabla_b v^{\mu a} = \del_b v^{\mu a}+{\itGamma^{\mu}}_{\nu\rho}
u^{\rho}_b v^{\nu a} + {\itGamma^a}_{cb} v^{\mu c}.
$$
Here, ${\itGamma^a}_{cb}$ is the induced connection on the world sheet. In
the absence of torsion, it is defined in terms of $\gamma_{ab}$ via the
known Christoffel formula. With this definition, the metricity condition is
satisfied for both metric tensors,
$$
\nabla_a \gamma_{bc} =0, \qquad \nabla_a g_{\mu\nu} =0 ,
$$
and (\ref{ResenaPrvaDekuplovanaJednacinaStrune}) is rewritten as
\begin{equation} \label{ResenaPrvaDekuplovanaJednacinaStruneKovarijantno}
\nabla_a (m^{ab}u^{\mu}_b) =0 .
\end{equation}
[The equivalent covariantization in the particle case could be achieved by
employing the one-dimensional induced connection $\itGamma
=(1/2\gamma)(d\gamma/d\tau)$.] Viewed as an equation for the string
trajectory, this equation contains the unknown coefficients $m^{ab}$. It can
be shown, however, that $m^{ab}$ are not fully arbitrary. Instead, they are
constrained by the same Eq.
(\ref{ResenaPrvaDekuplovanaJednacinaStruneKovarijantno}). To see this, we
project (\ref{ResenaPrvaDekuplovanaJednacinaStruneKovarijantno}) on
$u_{\mu}^c$, and obtain
\begin{equation} \label{JednacinaZaDokazDivergencijeM}
\nabla_a m^{ac}+m^{ab}u_{\mu}^c\nabla_a u^{\mu}_b = 0 .
\end{equation}
The second term is shown to identically vanish, and we end up with
\begin{equation} \label{KovarijantnoOdrzanjeM}
\nabla_a m^{ac} =0 .
\end{equation}
Thus, $m^{ab}$ is covariantly conserved, symmetric world sheet tensor. As
such, it is seen as the effective two-dimensional stress-energy tensor of
the string.

\subsection{\label{StringDiscussion}Discussion}

The Eq. (\ref{FinalBoundaryConditions}) represents a valid form of the
boundary conditions, irrespective of the world sheet parametrization used.
Even the situations when the coordinates $\xi^a$ are not well defined on the
boundary are included. If everywhere regular coordinates $\xi^a$ are used,
the coordinate vectors $u^{\mu}_a$ are linearly independent, and the
boundary conditions (\ref{FinalBoundaryConditions}) reduce to
\begin{equation} \label{ReseniGranicniUsloviStrune}
\sqrt{-\gamma}m^{ab}n_b\Big|_{\del\cM} =0 .
\end{equation}
This form of boundary conditions can further be simplified by employing the
standard parametrization $\xi^0=\tau$, $\xi^1=\sigma$. In these
coordinates, the boundary is defined by $\sigma =0$, $\sigma =\pi$,
the $n_0$ component of the normal (\ref{DefinicijaNormaleNaGranicu})
vanishes, and the boundary conditions take the simple form
\begin{equation} \label{ProstiGranicniUslovi}
\sqrt{-\gamma}m^{a1}\Big|_{\sigma = 0,\pi}=0 .
\end{equation}
Although the metric $\gamma_{ab}$ is assumed nondegenerate in the interior
of the world sheet, we shall retain the term $\sqrt{-\gamma}$ in the above
formula, to allow violations of this assumption on the boundary. This way,
we prevent loosing some important solutions of the world sheet equations. In
particular, the known Nambu-Goto dynamics belongs to this class of
solutions.

The boundary conditions obtained in this section are naturally associated with the
familiar Neumann boundary conditions of the conventional string theory. This is a
consequence of the fact that only "freely falling" strings in an external
gravitational field are considered. In the standard variational approach, they are
obtained by allowing free variation of the string boundary. The alternative choice is
to use Dirichlet boundary conditions, which are defined by imposing additional
constraints on the variation of the string boundary. Precisely, the string ends are
attached to an external $p$-brane, which (partially or fully) fixes their
trajectories.

In our approach, this situation is unsatisfactory, as the interaction of the string
with the $p$-brane violates the covariant conservation of the stress-energy tensor at
the string ends. We could, of course, impose these constraints by hand, but the
natural way to incorporate Dirichlet boundary conditions within our approach is to
consider the $p$-brane and the attached string as a single object moving in an external
gravitational field. Although such complex matter configurations are interesting by
themselves, we do not study them in the present work. Instead, we consider a string
that interacts only with the spacetime geometry. This means that the string ends have
nothing else to interact with, which in turn explains why the derivation of the
equations of motion yields precisely the Neumann boundary conditions.

The results of this section can be summarized as follows. We considered the
stress-energy conservation equation (\ref{ZOTEIuZgodnomObliku}), and looked
for a solution describing a stringlike distribution of matter. In the
lowest approximation (an infinitely thin string with no structure in the
transverse direction), the stress-energy tensor has the form
(\ref{StringCovariantTEI}), with $z^{\mu}(\xi)$ and $b^{\mu\nu}(\xi)$ the
unknown functions. We have found that (a) the string dynamics obeys the
equation
$$
\nabla_a (m^{ab}u^{\mu}_b) =0 ,
$$
and (b) the world sheet stress-energy tensor $m^{ab}$ is covariantly
conserved
$$
\nabla_a m^{ab} =0 .
$$
Further, (c) the end points of an open string are subject to the boundary
conditions
$$
\sqrt{-\gamma}m^{a1}\Big|_{\sigma =0,\pi} = 0 ,
$$
and (d) the target-space stress-energy tensor in this approximation has the
form
$$
\sqrt{-g}T^{\mu\nu}(x)=\int d^2\xi\sqrt{-\gamma}m^{ab}
u^{\mu}_a u^{\nu}_b\delta^{(4)}(x - z(\xi)).
$$

As opposed to the particle case, the dynamics of a stringy shaped matter
generally depends on its internal structure. Indeed, the two-dimensional
stress-energy conservation $\nabla_a m^{ab}=0$ has no unique solution. There
is a variety of possibilities to choose $m^{ab}$, each leading to a
different string dynamics.

Notice, however, that there exists a geometric solution analogous to the
one-dimensional $m\propto\gamma^{-1}$. It has the form
$m^{ab}\propto\gamma^{ab}$, and defines a string trajectory in full analogy
with the geodesic line. In fact, this particular choice of $m^{ab}$ yields
the string dynamics familiar from the literature: the obtained world sheet
equations and boundary conditions coincide with what we get by varying the
standard Nambu-Goto action. In the next section, we shall classify possible
canonical forms of $m^{ab}$, and explore their influence on the string
dynamics. We shall also give some examples to illustrate the feasibility of
stringlike solutions in ordinary field theories.

Let us note, in the end of this section, that the particle equations have
the same form as those of a string. Indeed, if the indices $a,b,\dots$ are
restrained to take only one value, say $0$, the world sheet equations will
get the form of a geodesic equation. This is not a coincidence. In fact, it
is possible to extend the whole discussion to a very general case of a
$p$-brane moving in a $D$-dimensional curved spacetime. The equations of
motion, boundary conditions, and the covariant conservation of the effective
mass tensor $m^{ab}$ are virtually the same, the only difference being in
the dimensionality of the world sheet. The $p$-brane world sheet coordinate
indices $a,b,\dots$ take the values $0,1,\dots,p$, while spacetime indices
$\mu,\nu,\dots$ range through $0,1,\dots, D-1$.

\section{\label{ExamplesSection}Internal structure of the string}

As we have seen in the previous section, the world sheet equations depend on
the type of matter the string is made of. To completely characterize
the string trajectory, we need the type and distribution of its mass tensor
$m^{ab}$. In this section, we shall classify possible canonical forms of
$m^{ab}$, and provide some illustrative examples.

\subsection{Canonical forms of the mass tensor}

In this section, we shall analyze the eigenproblem of the two-dimensional
mass tensor $m^{ab}$. The analogous $4$-dimensional analysis has been done
in \cite{Landau1975}, and the reduction to two dimensions is
straightforward.

The eigenproblem of $m^{ab}$ in a general world sheet with metric
$\gamma_{ab}$, is defined by the equation
$$
m^{ab}e_b=\lambda e^a ,
$$
where $e^a\equiv\gamma^{ab}e_b$. The existence of nonvanishing eigenvectors
$e^a$ is guaranteed by the condition $\det[m^{ab}-\lambda \gamma^{ab}]=0$.
It is rewritten as the quadratic equation
$$
\lambda^2-{m^a}_a\lambda+\gamma\det[m^{ab}]=0 ,
$$
with the discriminant
$$
\Delta\equiv\left({m^a}_a\right)^2-4\gamma\det[m^{ab}].
$$

Because of the indefiniteness of the metric, three cases are possible:
$\Delta>0$, $\Delta=0$, and $\Delta<0$. The eigenvectors can be either
timelike, spacelike or null. The mass tensor $m^{ab}$ cannot always be
diagonalized.

Let us analyze the behaviour of $m^{ab}$ in the vicinity of a point on the
world sheet. We shall use such $\xi^a$ coordinates which ensure
$\gamma_{ab}=\eta_{ab}$, and ${\itGamma^a}_{bc}=0$ in the chosen point.
If we write $m^{ab}$ in a matrix form as
$$
m^{ab} = \left(
\begin{array}{cc}
\rho & \pi \\
\pi & p \\
\end{array} \right) ,
$$
we see that $\rho$ represents energy density along the string, $\pi$ is the
energy flux, and $-p$ is the string tension. The components of the
stress-energy tensor are subject to the physical condition that energy flux
must not exceed the energy density: $\rho \geq |\pi|$. Otherwise, matter
would travel faster than light \cite{Landau1975}. This must be satisfied in
every reference frame, which can be shown to imply the general conditions on
the components of $m^{ab}$:
\begin{equation} \label{FizickiZahteviZaMateriju}
\rho + p\geq 2|\pi| , \qquad \rho\geq p .
\end{equation}

Now, we proceed to examine the cases where diagonalization is, or is not,
possible, and to give physical interpretation.

\subsubsection{Case $\Delta>0$}

In this case, one can employ a Lorentz transformation that brings
$m^{ab}$ to a diagonal form:
$$
m^{ab} = \left(
\begin{array}{cc}
\lambda^{(1)} & 0 \\
0 & -\lambda^{(2)} \\
\end{array} \right), \qquad \lambda^{(1)} \neq \lambda^{(2)} ,
$$
where $\lambda^{(1)}$ and $\lambda^{(2)}$ are the eigenvalues of $m^{ab}$.
This means that there exists a {\it rest frame}, where the energy flux is
zero, $\pi=0$, and matter does not move. This is the case for the usual
massive matter.

Conditions (\ref{FizickiZahteviZaMateriju}) are now rewritten as
$\lambda^{(1)}\geq|\lambda^{(2)}|$, which means that the energy density
$\rho$ is always positive, and exceeds the absolute value of the tension.
The string trajectory equations in the vicinity of the chosen world sheet
point can further be simplified by using a local inertial frame in the
target-space: $g_{\mu\nu}=\eta_{\mu\nu}$, ${\itGamma^{\mu}}_{\nu\rho}=0$.
Then, the world sheet equations $m^{ab}\nabla_a u^{\mu}_b=0$ reduce to
$$
\rho\frac{\del^2 z^{\mu}}{\del\tau^2}+
p\frac{\del^2 z^{\mu}}{\del\sigma^2} = 0 .
$$
If the string tension is positive, $p<0$, we may rewrite this as
\begin{equation} \label{WaveEquation}
\frac{\del^2 z^{\mu}}{\del\tau^2}-\omega^2
\frac{\del^2 z^{\mu}}{\del \sigma^2} = 0 ,
\end{equation}
where $\omega =\sqrt{-p/\rho}$ is the wave speed of the familiar wave
equation. The conditions (\ref{FizickiZahteviZaMateriju}) enforce $\rho >
-p$, wherefrom $\omega < 1$. Thus, the speed of sound along the string is
less than that of light, as expected for ordinary massive matter.

The world sheet metric $\gamma_{ab}$ is assumed to be everywhere
nondegenerate, including the boundary itself. In this case, the boundary
conditions (\ref{ReseniGranicniUsloviStrune}) reduce to
\begin{equation} \label{GranicniUsloviZgodniZaAnalizuMaterije}
\lambda^{(1)}n^0\Big|_{\del\cM}=
\lambda^{(2)}n^1\Big|_{\del\cM}= 0 ,
\end{equation}
which means that at least one of the eigenvalues must vanish. The physical
condition $\lambda^{(1)}\geq |\lambda^{(2)}|$ then singles out
$\lambda^{(1)}\neq 0$, $\lambda^{(2)}=0$, and consequently,
$$
m^{ab}\Big|_{\del\cM} =
\left(
\begin{array}{cc}
\rho & 0 \\
0 & 0 \\
\end{array}
\right) ,
$$
with the interpretation that the tension $-p$ vanishes on the string ends.
The conditions (\ref{GranicniUsloviZgodniZaAnalizuMaterije}) also imply
$n^0=0$, which means that the boundary coincides with a coordinate line
$\sigma=\text{const.}$. Thus, the form (\ref{ProstiGranicniUslovi}) of the
boundary conditions could have been used, too.

\subsubsection{Case $\Delta=0$}

In this case, there exists a boost that brings $m^{ab}$ to the form
$$
m^{ab} = \left(
\begin{array}{cc}
\lambda + \mu & \mu \\
\mu & -\lambda + \mu \\
\end{array} \right).
$$
Here, $\lambda$ is the single eigenvalue, and the Lorentz invariant sign of
$\mu$ defines three subcases: $\mu>0$, $\mu=0$ and $\mu<0$. The conditions
(\ref{FizickiZahteviZaMateriju}) reduce to $\lambda \geq 0$ and $\mu\geq 0$,
which excludes the third possibility $\mu<0$ as nonphysical. Thus, every
nontrivial $m^{ab}$ is the sum of matrices corresponding to the cases
$\lambda=0$, $\mu>0$ and $\lambda>0$, $\mu=0$. Let us discuss these two
situations.

In the case $\bm{\lambda=0}$, $\bm{\mu>0}$, the only eigenvector of
$m^{ab}$ is lightlike, and no rest frame exists. The situation is
interpreted as that of a {\it massless matter}. In the four-dimensional
electrodynamics, for example, we can consider electric and magnetic fields
of equal intensity, $E=B$, and perpendicular to each other, $\vec E\cdot\vec
B =0$. In a suitable reference frame, the stress-energy tensor has the form
$$
T^{\mu\nu} = \left(
\begin{array}{cccc}
E^2 &  E^2 & 0 & 0 \\
E^2 &  E^2 & 0 & 0 \\
0 & 0 & 0 & 0 \\
0 & 0 & 0 & 0 \\
\end{array}
\right).
$$
Such is, for example, linearly polarized plane wave propagating in the $x$
direction. Its stress-energy tensor belongs to the class under
consideration, with $\mu=E^2$.

The boundary conditions (\ref{ReseniGranicniUsloviStrune}) reduce to
$$
\mu (n^0-n^1)\Big|_{\del\cM}=0,
$$
wherefrom $n^0=n^1$. Thus, the normal to the boundary is lightlike, and the
definition (\ref{DefinicijaNormaleNaGranicu}) implies the same for the
boundary itself. We see that the boundary cannot coincide with the
coordinate line $\sigma=\text{const.}$, which is the reason we could not use
the boundary conditions in the form (\ref{ProstiGranicniUslovi}).

The case $\bm{\lambda>0}$, $\bm{\mu=0}$. Here, the
eigenvalue $\lambda$ is degenerate, and the two eigenvectors can be chosen
to be spacelike and timelike, respectively. The mass tensor is not only
diagonal, but proportional to the metric: $m^{ab}=\lambda\eta^{ab}$. This
can covariantly be written as $m^{ab}=\lambda\gamma^{ab}$, and defines the
known {\it Nambu-Goto string}. The energy density $\rho$ is positive, and
equal to the tension:
$$
\rho = -p .
$$
The equations of motion are precisely the minimal-surface equations. In a
local inertial frame, they reduce to the wave equation
$$
\frac{\del^2 z^{\mu}}{\del\tau^2}-\frac{\del^2z^{\mu}}{\del\sigma^2}=0 ,
$$
which is a special case of (\ref{WaveEquation}), with $\omega=1$. Thus, the
speed of sound in the string equals the speed of light. No conventional
elastic material exhibits such a behaviour.

In the local inertial frame ($\gamma_{ab}=\eta_{ab}$), the only appropriate
form of boundary conditions is (\ref{FinalBoundaryConditions}). Indeed, it
reduces to
$$
n^a u^{\mu}_a \Big|_{\del\cM}=0,
$$
which shows that the coordinate vectors $u^{\mu}_a$ are not linearly
independent. Thus, the inertial frame is necessarily degenerate at the
boundary. The best we can do is to Lorentz rotate the normal $n^a$ to
achieve $n^0=0$, and bring the boundary conditions to the conventional form
$u^{\mu}_1=0$. If, on the other hand, we insist on using regular
parametrization of the world sheet, we must leave the inertial frame. Only
then, we can use the form (\ref{ProstiGranicniUslovi}) of the boundary
conditions, and obtain
\begin{equation} \label{NambuGotoBoundaryConditions}
\sqrt{-\gamma}\gamma^{a1}\Big|_{\sigma=0,\pi} = 0.
\end{equation}
We see that the world-sheet metric at the boundary is degenerate. A careful
analysis of the conditions (\ref{NambuGotoBoundaryConditions}) yields the
solution
$$
\gamma_{ab}=
\left(
\begin{array}{cc}
0 & 0 \\
0 & \gamma_{11} \\
\end{array}
\right)
$$
at $\sigma=0$ and $\sigma=\pi$. In particular, $\gamma_{00}\equiv g_{\mu\nu}u^{\mu}_0
u^{\nu}_0=0$, which means that the world sheet boundary is lightlike. This is the
familiar result of the Nambu-Goto dynamics: the string ends move with the speed of
light.

The boundary conditions we have derived are the familiar Neumann boundary conditions.
They are the consequence of the "free falling" character of the string motion, and
are obtained automatically in our approach. In contrast, the Dirichlet boundary
conditions demand the string ends to be attached to an external $p$-brane, as
explained in Sec. \ref{StringDiscussion}.

In both $\Delta=0$ cases we have considered, the string ends move with the
speed of light. There is a difference, however, in the behaviour of the
induced metric $\gamma_{ab}$. In the massless case, the metric is regular at
the boundary ($\gamma\neq 0$), while in the Nambu-Goto case, it is
degenerate ($\gamma=0$). In geometric terms, the world sheet either
intersects the target-space light cone, or just touches it. Both
world sheets are regular $2$-dimensional surfaces, though.

\subsubsection{Case $\Delta<0$}

In this case, there exists a boost that brings $m^{ab}$ to the form
$$
m^{ab} = \left(
\begin{array}{cc}
\lambda'  & \lambda'' \\
\lambda'' & -\lambda' \\
\end{array} \right).
$$
Here, the two eigenvalues are complex-conjugate, $\lambda^{(0)}=\lambda'
-i\lambda''$ and $\lambda^{(1)}=\lambda' +i\lambda''$. The corresponding
eigenvectors are also complex.

The conditions (\ref{FizickiZahteviZaMateriju}) are in contradiction with
the above form of $m^{ab}$. This means that they are never satisfied, as one
can always find a reference frame where energy flux exceeds the energy
density. Thus, the case is unphysical, corresponding to matter whose speed
exceeds the speed of light.

To summarize, the eigenvalue problem we have considered brought about three
possible cases: a) the case $\Delta>0$ describes massive matter, b) the case
$\Delta=0$ combines massless matter with matter of Nambu-Goto type,
and c) the case $\Delta<0$ represents unphysical, tachyonic matter.

\subsection{Inhomogeneous example}

Let us consider a string characterized by a highly inhomogeneous
distribution of matter. Take the radical situation when all the mass is
localized in one single point. In the lowest approximation, the
mass tensor $m^{ab}$ is chosen in the form
\begin{equation} \label{MaseniTenzorLinijeNaStruni}
\sqrt{-\gamma}m^{ab}=\int ds b^{ab}(s)\delta^{(2)}(\xi-\chi(s)),
\end{equation}
where $b^{ab}(s)$ are some parameters, and $\xi^a=\chi^a(s)$ is a line on
the world sheet, parametrized by the proper distance $s$. As a
consequence, the corresponding tangent vector $v^a\equiv d\chi^a/ds$ has
unit norm: $\gamma_{ab}v^av^b=1$. The world line equation is expected to be
found by the analysis of the corresponding world sheet equations.

Our strategy is the same as that of Sec.\ \ref{ParticleSection}, the only
difference being the dimensionality of the target-space (there $4$, here
$2$). Therefore, we start with the conservation law $\nabla_a m^{ab}=0$, and
apply the ansatz (\ref{MaseniTenzorLinijeNaStruni}). The result is the exact
analogue of the $4$-dimensional case: $b^{ab}\propto v^av^b$, and the
world line is a world sheet geodesic
\begin{equation} \label{JednacineKretanjaLinijeNaStruni}
\frac{d v^a}{ds}+{\itGamma^{a}}_{bc}v^b v^c = 0.
\end{equation}
The question is if this particular world sheet geodesic is also a spacetime
geodesic, as one would expect. The answer is affirmative. To see this, we
first note that the world line tangent vector $\vec v$ is also tangent to
the world sheet and the spacetime. Its spacetime components $v^{\mu}$ can be
expressed in terms of the world sheet components $v^a$ as follows:
$$
v^{\mu} = \frac{dz^{\mu}(\chi(s))}{ds} =
\frac{\del z^{\mu}(\chi)}{\del\chi^a}
\frac{d\chi^a(s)}{ds} = u^{\mu}_a v^a .
$$
The spacetime proper distance $s$ is, of course, the same as the world sheet
proper distance. To check if our line is a spacetime geodesic, we use
$v^{\mu}=u^{\mu}_av^a$ and (\ref{JednacineKretanjaLinijeNaStruni}) to find
that
$$
\frac{d v^{\mu}}{d\tau} + v^{\rho}v^{\sigma}
{\itGamma^{\mu}}_{\rho\sigma} = v^av^b \nabla_au^{\mu}_b .
$$
Now, the world sheet equations $m^{ab}\nabla_au^{\mu}_b=0$, and the
fact that $m^{ab}$, when calculated on the line $\xi^a=\chi^a(s)$, is
proportional to $v^av^b$ lead us to
$$
\frac{d v^{\mu}}{ds}+{\itGamma^{\mu}}_{\rho\sigma}
v^{\rho}v^{\sigma} = 0 .
$$
This is a spacetime geodesic equation, as we expected.

\subsection{Nielsen-Olesen vortex line}

In our second example, we shall evaluate the stress-energy tensor of the
known Nielsen-Olesen vortex line field configuration \cite{Nielsen1973}. We
start with the Higgs type of Lagrangian in Minkowski spacetime:
$$
{\cal L} = -\frac{1}{4}F_{\mu\nu}F^{\mu\nu}
+\frac{1}{2}(\nabla_{\mu}\phi)(\nabla^{\mu}\phi)^\ast
-\lambda (\phi\phi^\ast - a^2)^2 ,
$$
where $\nabla_{\mu}\phi\equiv (\partial_{\mu}+ieA_{\mu})\phi$. It was
shown in \cite{Nielsen1973} that the corresponding field equations allow for
a static, axially symmetric solution localized around the $z$-axis. In polar
coordinates ($x=\rho\cos\varphi$, $y=\rho\sin\varphi$), and the time gauge
$A^0=0$, the solution has the form
$$
\vec A = A\vec e_{\varphi},\quad
\phi=|\phi|e^{-i\varphi},
$$
where $A$ and $|\phi|$ are functions of $\rho$ only, and $\vec e_{\varphi}$
stands for the unit vector in the $\varphi$ direction. The unknown functions
$A(\rho)$ and $|\phi|(\rho)$ are determined by the field equations. Far from
the vortex core ($z$-axis), the Nielsen-Olesen solution rapidly approaches
the vacuum
\begin{equation} \label{TrueVacuum}
A=-\frac{1}{e\rho},\quad |\phi|=a .
\end{equation}
It is characterized by the absence of electromagnetic field, $\vec B =\vec E
=0$, and represents the true vacuum of the theory. In the core, when
$\rho\to 0$, the solution approaches the false vacuum
\begin{equation} \label{FalseVacuum}
A=\frac{B}{2}\rho ,\quad |\phi|=0 ,
\end{equation}
with constant magnetic field $\vec B =B\vec e_z$, and vanishing electric
field $\vec E =0$.

As we do not know the exact analytic form of the Nielsen-Olesen vortex line
solution, we shall approximately view it as follows. Let us define the
parameter $\ell$ to measure the vortex width. In the region
$0\leq\rho\leq\ell$, we assume the false vacuum solution
(\ref{FalseVacuum}), while outside that region, the solution is taken to be
that of the true vacuum (\ref{TrueVacuum}):
\begin{eqnarray*}
& 0\leq\rho\leq\ell : &  \quad A=\frac{B}{2}\rho ,
                         \enskip |\phi|=0 ,         \\
& \rho>\ell :         &  \quad A=-\frac{1}{e\rho} ,
                         \enskip |\phi|=a .
\end{eqnarray*}
The continuity of the function $A(\rho)$ requires the magnetic field to
satisfy $B=-2/e\ell^2$. In the limit $\ell\to 0$ (stringlike solution),
$B\to\infty$, but the magnetic flux retains the constant value
$-2\pi/e$.

Now, we are ready to evaluate the stress-energy tensor
$$
T_{\mu\nu}=
F_{\mu\lambda}F^{\lambda}{}_{\nu} +
(\nabla_{(\mu}\phi)(\nabla_{\nu)}\phi)^\ast -
\eta_{\mu\nu}{\cal L} ,
$$
where indices in round brackets are symmetrized. A simple calculation shows
that it vanishes outside the vortex core, $T^{\mu\nu}=0$ if $\rho>\ell$,
while in the core, $0\leq\rho\leq\ell$, it has diagonal form with
\begin{eqnarray*}
&& T^{00}=-T^{33}=\frac{2}{e^2\ell^4}+\lambda a^4 , \\
&& T^{11}=T^{22}=\frac{2}{e^2\ell^4}-\lambda a^4 .
\end{eqnarray*}
In the limit $\ell\to 0$, the vortex solution looks like an infinite
string whose world sheet coincides with the \mbox{$t$--$z$} plane. Using the
parametrization $\xi^0=t$, $\xi^1=z$, the world sheet coordinate vectors
become $u^{\mu}_0=\delta^{\mu}_0$, $u^{\mu}_1=\delta^{\mu}_3$, and the
induced metric $\gamma_{ab}$ reduces to $\eta_{ab}$. We see that the
stress-energy tensor can be written in the form (\ref{StringCovariantTEI})
with
$$
b^{\mu\nu}=\pi\ell^2 T^{\mu\nu}.
$$
To obtain a valid string solution, we must get rid of the undesirable
tension in the transverse direction. To this end, we adjust our free
parameters to obey the constraint
\begin{equation} \label{Constraint}
\lambda a^4=\frac{2}{e^2\ell^4} ,
\end{equation}
whereupon $T^{00}=-T^{33}=4/e^2\ell^4$ remain the only nonvanishing
components of the stress-energy tensor. Now, our $b^{\mu\nu}$ takes
the form (\ref{ResenaDrugaDekuplovanaJednacinaStrune}) with
\begin{equation} \label{VortexTension}
 m^{ab}=m\,\eta^{ab},\quad  m\equiv\frac{4\pi}{e^2\ell^2}.
\end{equation}
This is exactly the form of $m^{ab}$ that defines Nambu-Goto string. It is
easy to check then that $t$-$z$ plane satisfies the world sheet equations
(\ref{ResenaPrvaDekuplovanaJednacinaStruneKovarijantno}).

Let us note in the end, that our approximation is in good agreement with the
analysis of Nielsen and Olesen \cite{Nielsen1973}. They found the range of
parameters that enables their vortex solution to be viewed as Nambu-Goto
string. In our notation, $\lambda\sim e^2\sim (a\ell)^{-2}\gg 1$, and
$m\sim a^2$. This agrees with both our constraint (\ref{Constraint}) and
our Eq. (\ref{VortexTension}). In particular, we see that the coupling
constant $e$ must be very large (of the order $\ell^{-1}$) to ensure finite
tension in the limit $\ell\to 0$.

\section{\label{Conclusions}Concluding remarks}

The analysis in this paper concerns the dynamics of realistic material
strings in curved backgrounds. In the simple case we have considered, the
background geometry is Riemannian, defined in terms of the metric
tensor alone. The dynamics of geometry and matter fields is governed by the
Einstein's equations.

In the specific setting considered, we assume the existence of a stable
stringlike kink configuration. The type of matter fields involved is not
specified. We only assume that matter fields are sharply localized around a
line, while geometry itself is not. For practical reasons, the matter fields
are considered weak enough to have negligible influence on the background
geometry. This way, the target-space metric is attributed the properties of
an external field, insensitive to string dynamics.

The method used is, basically, the Mathisson-Pa\-pa\-pe\-t\-rou method for
pointlike matter \cite{Mathison,Papapetrou1951} generalized to linelike
configurations. We make use of the fact that every exponentially decreasing
function can be written as a series of derivatives of Dirac $\delta$
function. The Mathisson-Papapetrou multipole moments are then obtained as
the coefficients in the expansion. We use this method to expand the
covariantly conserved stress-energy tensor of matter fields. The
world sheet equations are obtained in the lowest order --- the approximation
of an infinitely thin string.

The results of our analysis can be summarized as follows. The dynamics of a
stringy shaped matter in torsionless spacetimes generally depends on the
internal structure of the string. The coefficients $m^{ab}$ entering the
world sheet equations are the components of the covariantly conserved
effective $2$-dimensional stress-energy tensor of the string. As opposed
to the point particle case, $m^{ab}$ can not generally be eliminated by
world sheet reparametrizations. The diversity of possible forms of $m^{ab}$
has been analyzed with the emphasis on two questions. The first is the
question of homogeneity of matter distribution along the string. The second
is the classification of possible canonical forms of $m^{ab}$.

The possibility of unevenly distributed matter was demonstrated in an
extreme case. If all the matter is localized in one point on the string, it
was shown that the world sheet equations boil down to the geodesic equation,
as expected.

We have also examined the possible canonical forms of $m^{ab}$. The most
interesting is the case of homogeneously distributed matter whose tension
equals its mass density. In this case, the known Nambu-Goto string dynamics
is discovered. To demonstrate that such kink configurations are indeed
possible, the Nielsen-Olesen vortex line solution of a Higgs type scalar
electrodynamics is given as an example.

Before we close our exposition, let us mention that our main result can
easily be generalized to include arbitrary $p$-brane distribution of matter.
The corresponding world-sheet equations are of the same form
$$
\nabla_a\left(m^{ab}u^{\mu}_b\right)=0 ,
$$
but this time $a,b = 0,1,\dots ,p$, and $m^{ab}$ is the covariantly
conserved $(p+1)$-dimensional energy-momentum tensor of the brane.
Obviously, the diversity of possible forms of $m^{ab}$ is bigger than in the
string case. The known minimal surface equations are obtained for $m^{ab}
\propto\gamma^{ab}$, where $\gamma_{ab}$ is the induced world sheet metric.

Let us say in the end that these are just preliminary results before we
address the more important problem of string dynamics in general backgrounds
with torsion.

\begin{acknowledgments}
This work was supported by the Serbian Science Foundation, Serbia.
\end{acknowledgments}

\appendix*

\section{Series expansion in $\bm{\delta}$ function derivatives}

Here, we develop a formalism to expand a given function into a series of
derivatives of Dirac $\delta$ function. After that, we give an intuitive
interpretation of the result, and associate it to the well-known multipole
expansion in electrodynamics.

Consider a real valued function $f(x)$, and write its Fourier integral
$$
f(x) = \int dk \tilde{f}(k) e^{ikx}.
$$
We can expand $\tilde{f}(k)$ into the power series around $k=0$,
$$
\tilde{f}(k) = \sum_{n=0}^{\infty} a_n k^n ,
$$
and rewrite the function $f(x)$ as
$$
f(x) = \sum_{n=0}^{\infty} a_n  \int dk k^n e^{ikx}.
$$
The integral on the right-hand side is evaluated by means of the identity
$$
\int dk e^{ikx} = 2\pi \delta(x),
$$
which gives
$$
\int dk k^n e^{ikx} = 2\pi (-i)^n \frac{d^n}{dx^n} \delta (x).
$$
Therefore, the function $f(x)$ can be expanded into an infinite
series of derivatives of Dirac $\delta$ function as follows:
$$
f(x) = \sum_{n=0}^{\infty} b_n \frac{d^n}{dx^n} \delta (x).
$$
The coefficients $b_n$ are given by
$$
b_n = \frac{(-1)^n}{n!} \int dx  x^n f(x),
$$
and are usually called $n$-th order moments of the function $f(x)$.
The coefficients $b_n$ are well defined if the function $f(x)$ decreases
faster than any power of $x$. In particular, the exponentially decreasing
function has well defined $\delta$ function expansion.

The above procedure can easily be extended to include a higher dimensional
case. Given a function $f(x) \equiv f(x_1,\dots,x_d)$, and a point $z\equiv
(z_1,\dots,z_d)$, one can expand $f(x)$ via the general formula
\begin{equation} \label{app:RazvojFunkcijePoDeltaFunkciji}
f(x) = \sum_{n=0}^{\infty} b^{\mu_1\dots \mu_n} \del_{\mu_1} \dots
\del_{\mu_n} \delta^{(d)}(x-z).
\end{equation}
Here, $\delta^{(d)}(x-z)$ is a $d$-dimensional $\delta$ function,
$\del_{\mu}\equiv \del/\del x^{\mu}$, and the indices $\mu_i$ take values
from $1$ to $d$. The corresponding formula for the coefficients reads:
\begin{equation} \label{app:KoeficijentiRazvojaFunkcijePoDeltaFunkciji}
b^{\mu_1\dots \mu_n} = \frac{(-1)^n}{n!}
\int d^dx f(x)\prod_{i=1}^n (x^{\mu_i}-z^{\mu_i}) .
\end{equation}

The intuitive interpretation of the expansion
(\ref{app:RazvojFunkcijePoDeltaFunkciji}) goes as follows. Suppose a
function $f(x)$ is localized around the point $z$, and is rapidly
approaching zero as one moves away from $z$. If we observe the function from
a distance, we can approximate it with the $\delta$ function, which is the
first term in (\ref{app:RazvojFunkcijePoDeltaFunkciji}). As we get closer to
$z$, we see more "structure" in $f(x)$. In the formalism, this is described
by higher order terms in (\ref{app:RazvojFunkcijePoDeltaFunkciji}). The
better localized the function $f(x)$, the less significant is the
contribution of higher order terms.

As an example, consider electrostatic charge density $\rho(\vec{x})$ of a
localized source. Expanding it with respect to $\vec{x}=0$, one gets the
first two coefficients:
\begin{eqnarray*}
n=0: && b=\int d^3x \rho(\vec{x})=Q ,        \\
n=1: && \vec{b}=-\int d^3x \vec{x}\rho(\vec{x})=-\vec{p} .
\end{eqnarray*}
We recognize $Q$ and $\vec{p}$ as the total charge and the electrostatic
dipole moment of the source. So, we can write the expansion as
$$
\rho(\vec{x})=Q\delta^{(3)}(\vec{x})-
\vec{p}\cdot\vec\nabla\delta^{(3)}(\vec{x})+\cdots .
$$
The electrostatic potential $\varphi(\vec{x})$ is calculated from
$$
\varphi(\vec{x})=\int d^3y \frac{\rho(\vec{y})}{|\vec{x} - \vec{y}|},
$$
whereupon the well-known multipole expansion in electrodynamics is obtained
\cite{Jackson1974}:
$$
\varphi(\vec{x})=\frac{Q}{r}+\frac{\vec{p}\cdot\vec{x}}{r^3}+\cdots .
$$

This example illustrates the use of the $\delta$ function expansion, and
clarifies what type of approximation is done when one ignores all but the
$n=0$ term in the expansion.

\bigskip
\textbf{Note added}
\bigskip

After this paper was published, the authors became aware of another paper
dealing with the same topic \cite{Deser1975}, albeit the analysis given
there seems somewhat incomplete.

\end{document}